# PREVENTING SQL INJECTION ATTACK USING PATTERN MATCHING ALGORITHYM

Swapnil Kharche[1] , Jagdish patil[2]

Kanchan Gohad[3] , Bharti Ambetkar[4]

Department of Computer Engineering

MAHARASHTRA INSTITUTE OF TECHNOLGY

PUNE

*kharchessk@gmail.com[1]* , *jagspatil91@gmail.com[2]*

*Ksg.gohad@gmail.com[3]* , *ambetkarbharati34@gmail.com[4]*

**Abstract:**

SQL injection attacks, a class of injection flaw in which specially crafted input strings leads to illegal queries to databases, are one of the topmost threats to web applications. A Number of research prototypes and commercial products that maintain the queries structure in web applications have been developed. But these techniques either fail to address the full scope of the problem or have limitations. Based on our observation that the injected string in a SQL injection attack is interpreted differently on different databases. A characteristic diagnostic feature of SQL injection attacks is that they change the intended structure of queries issued. Pattern matching is a technique that can be used to identify or detect any anomaly packet from a sequential action. Injection attack is a method that can inject any kind of malicious string or anomaly string on the original string. Most of the pattern based techniques are used static analysis and patterns are generated from the attacked statements. In this paper, we proposed a detection and prevention technique for preventing SQL Injection Attack (SQLIA) using Aho–Corasick pattern matching algorithm. In this paper, we proposed an overview of the architecture. In the initial stage evaluation, we consider some sample of standard attack patterns and it shows that the proposed algorithm is works well against the SQL Injection Attack.

# I. INTRODUCTION

Companies and organizations use web applications to provide better service to the end users. The Databases used in web applications often contain confidential and personal Information SQL Injection attacks target databases that are accessible through a web front-end, and take advantage of flaws in the input validation logic of Web components such as CGI scripts. This communication is commonly done through a low– level API by dynamically constructing query strings with in a general purpose programming language. This low–level interaction (or) communication is dynamic (or) session based because it does not take into account the structure of the output language. The user input statements are treated as isolated lexical entries (or) string. Any attacker can embed a command in this string, which poses a serious threat to web application security. In the last few months application-level vulnerabilities have been exploited with serious consequences by the hackers have tricked e-commerce sites into shipping goods for no charge, usernames and passwords have been harvested and confidential information such as addresses and credit-card numbers has been leaked. The reason for this occurrence is that web applications and detection systems do not know the attacks thoroughly and use limited sets of attack patterns during evaluation. SQL Injection attacks can be easily prevented by applying more secure authentication schemes in login phase itself. Most of the attacks made on the web target the vulnerability of web applications. Vulnerabilities of web ability to obtain and charge the sensitive information, such as military systems, banks, and e-business, etc are exposed to a great security risk. SQL Injection Attack (SQLIA) is one of the very serious threats for web applications. Pattern matching is a technique that can be used to identify or detect any anomaly packet from a sequential action. Injection attack is a method that can inject any kind of malicious string or anomaly string on the original string.

For better understanding let us have look at the following example. We all know that most of the applications that we are accessing through internet will have a login page to authenticate the user who is using the application. Figure 1 show such a login page. Here when a user is submitting his username and password, an SQL query is generated in the back end to check whether the given credentials are valid or not. Suppose the given username is 1 and password is 111, the query will be:

*Select * from login where user='1' and pass='111'*

This is the normal case and if any rows are selected by the query, the user is allowed to log in. Now, will see attacker scenario That is an attacker wants to log in without correct username and password. Instead of entering valid username if he uses injection string like "hacker' OR '1'='1'—" as username and "something" as password, the query formed will be like this:

*Select * from login where user='hacker' or '1'='1' –' and pass='something'*

When this query is executed in the database, it will always return a true and the authentication will succeed.

**II. RELATED WORK**

Although web application can be classified as programs running on a web browser, web applications generally have a Tree-tier construction
1) Presentation Tier: receives the user's input data and shows the result of the processed data to the user. It can be thought of as the Graphic User Interface (GUI). Flash, HTML, Javascript, etc. are Directly interact with the user.

2) CGI Tier: also known as the Server Script Process, is located in between the presentation tier and database tier. The data inputted by the user is processed and stored into the database. The database sends back the stored data to the CGI tier which is finally sent to the presentation tier for viewing. Therefore, the data processing within the web application is done at the CGI Tier. It can be programmed in various server script languages such as JSP, PHP, ASP, etc. 3) Database Tier: stores and manages all of the processed user's input data. All sensitive data of web applications are stored and managed within the database.

The database tier is responsible for the access of authentication users. A general framework for detecting malicious database transaction patterns using data mining was proposed by Bertino et al [1 2] to mine database logs to form user profiles that can model normal behaviors and identify anomalous transaction in database with role based access control mechanism. The system is able to identify intruder by detecting behaviors that different from the normal behavior. Kamra et al [3], proposed an enhanced model that can identify intruders in databases where there are no roles associated with each user. Bertino et al [4], proposed a framework based on anomaly detection technique and association rule mining to identify the query that deviates from the normal database application behavior. Bandhakavi et al [5] proposed a misuse detection technique to detect SQLIA by discovering the intent of a query dynamically and then comparing the structure of the identified query with normal queries based on the user input with the discovered intent.

The contribution of this paper is to propose a technique for detecting and preventing SQLIA using both static phase and dynamic phase. The proposed technique uses static Anomaly Detection using Aho–Corasick Pattern matching algorithm. The anomaly SQL Queries are detection in static phase. In the dynamic phase, if any of the query is identified as anomaly query then new pattern will be created from the SQL Query and it will be added to the Static Pattern List (SPL).

### III. PROPOSED SCHEME

In this section, we introduce an efficient algorithm for detecting and preventing SQL Injection Attack using Aho–Corasick Pattern matching algorithm. The proposed architecture is given in figure 1 below. The proposed scheme has the following two modules, 1) Static Phase and 2) Dynamic Phase In the Static Pattern List, we maintain a list of known Anomaly Pattern. In Static Phase, the user generated SQL Queries are checked by applying Static Pattern Matching Algorithm. In Dynamic Phase, if any form of new anomaly is occur then Alarm will indicate and new Anomaly Pattern will be generated. The new anomaly pattern will be updated to the Static Pattern List. The following steps are performed during Static and Dynamic Phase,

**Static Phase**

Step 1: User generated SQL Query is send to the proposed Static Pattern Matching Algorithm

Step 2: The Static Pattern Matching Algorithm is given in Pseudo Code is given below

Step 3: The Anomaly patterns are maintained in Static Pattern List, during the pattern matching process each pattern is compared with the stored Anomaly Pattern in the list

Step 4: If the pattern is exactly match with one of the stored pattern in the Anomaly Pattern List then the SQL Query is affected with SQL Injection Attack

**Dynamic Phase**

Step 1: Otherwise, Anomaly Score value is calculated for the user generated SQL Query, If the Anomaly Score value is more then the Threshold value, then a Alarm is given and Query will be pass to the Administrator.

Step 2: If the Administrator receives any Alarm then the Query will be analyze by manually. If the query is affected by any type of injection attack then a pattern will be generated and the pattern will be added to the Static Pattern list

*A. Anomaly Score value calculation*

In the static phase, each anomaly pattern from the Static Pattern List is checked with the user generated query. The proposed scheme will calculate the Anomaly Score value of the query for each pattern in the Static Pattern List. If the Query is match 100% with any of the pattern from the Static Pattern List, then Query is affected with SQL Injection Attack. Otherwise, the high matching score is called as an Anomaly Score value of a query. If the Anomaly Score value is more then the Threshold value (assume that 50%), then the query will be transfer to the Administrator.

*B. Pattern Matching Algorithm*

In the proposed architecture, Static Pattern Matching Algorithm is the main part and the pseudo code for the Algorithm is given below,

```
Static Pattern Matching Algorithm
1:  Procedure SPMA(Query, SPL[ ])
    INPUT: Query← User Generated Query
           SPL[ ]←Static Pattern List with m Anomaly
                  Pattern
2:  For j = 1 to m do
3:  If (AC (Query, String.Length(Query), SPL[j][0]) == φ)
    then
4:      Anomaly_score = (Matching_value(Query, SPL[j]) / StringLength(SPL[j])) × 100
5:      If (Anomaly_Score ≥ Threshold_Value)
6:      then
7:          Return Alarm → Administrator
8:      Else
9:          Return Query → Accepted
10:     End If
11: Else
12:     Return Query → Rejected
13: End If
14: End For
End Procedure
```

### C. Aho–Corasick Algorithm

There are many approaches to recognizing patterns that involve using finite automata. The Aho–Corasick algorithm is one such classic algorithm. The idea is that a finite automaton is constructed using the set of keywords during the pre–computation phase of the algorithm and the matching involves the automaton scanning the SQL query statement reading every character in SQL query exactly once and taking constant time for each read of a character. Pseudo code of the Aho–Corasick multiple key word matching algorithm is given below,

```
Aho – Corasick Multiple Keyword Matching Algorithm
1:  Procedure AC(y,n,q0)
    INPUT: y← array of m bytes representing the text input
           (SQL Query Statement)
           n← integer representing the text length
           (SQL Query Length)
           q0←initial state (first character in pattern)
2:  State ← q0
3:  For i = 1 to n do
4:      While g( State, y[i] == fail) do
5:          State ← f(State)
6:      End While
7:      State ← g(State, y[i])
8:      If o(State) ≠ φ then
9:          Output i
10:     Else
11:         Output ψ
12:     End If
13: End for
14: End Procedure
```

The AC algorithm uses a refinement of a tries to store the set of Anomaly Keywords in a pattern matching *Example* The sample of Query generation is given in figure 2 and the process of pattern matching for the user generated query in given in Figure 3 and Figure 4

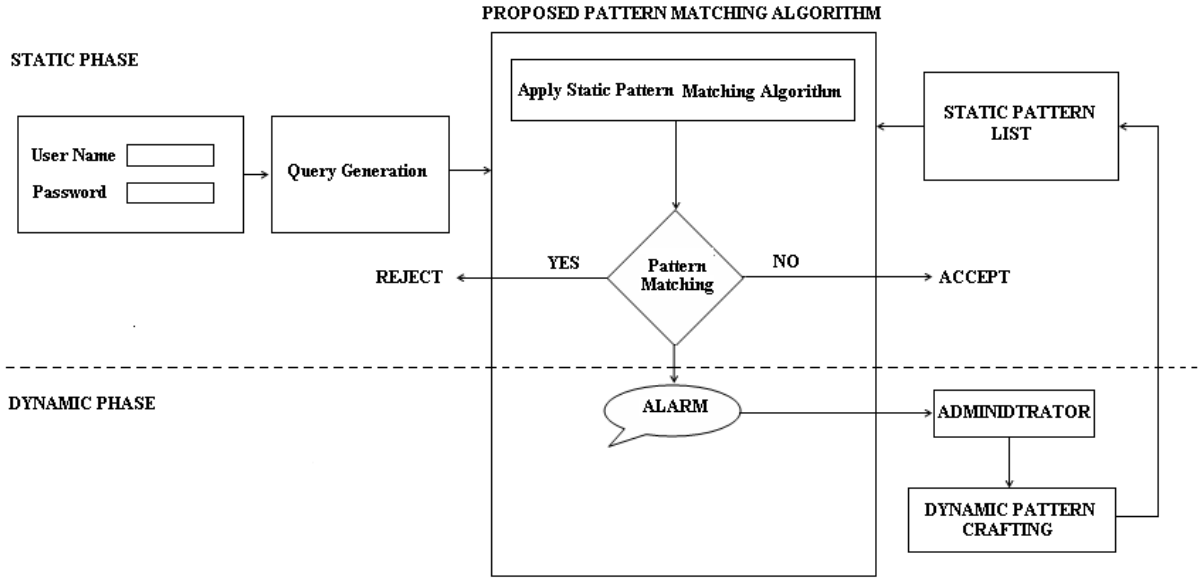

Fig. 1. Proposed Architecture

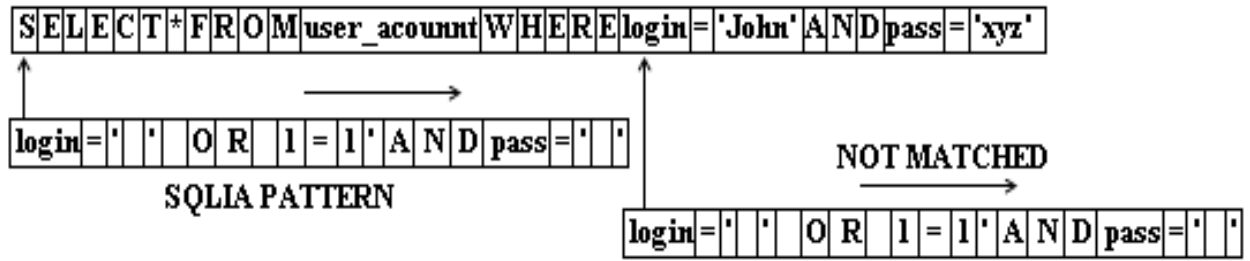

Fig. 2. SQL Query Generation with legal user name and password

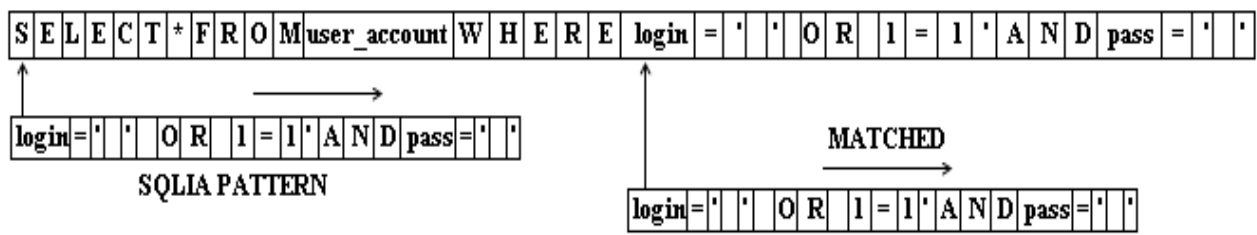

Fig. 3. SQLIA Pattern Matching Process

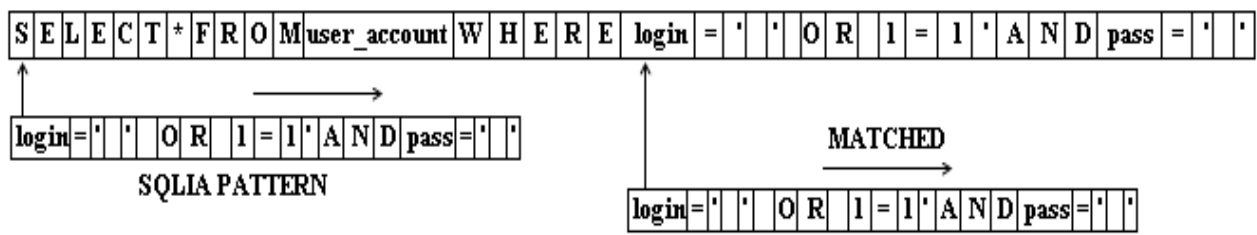

Fig. 4. SQLIA Pattern Exactly Matching

## IV. CONCLUSION

This paper describes the challenges that Internet applications that make use of a database system face in terms of security and protection the private data. SQL-Injection attacks is a legitimate threat that endangers the confidentiality of data and may cost an organization a great deal of money and even their reputation for different tools. The system is significantly safer from both online threats. there will be constant need to protect it. So we have proposed a scheme for detection and prevention of SQL Injection Attack using Aho–Corasick pattern matching algorithm. The proposed scheme is evaluated by using sample of well known attack patterns. Initial stage evaluation shows that the proposed scheme is produce not false positive and false negative. The pattern matching process takes minimum of O (n) time